\documentclass[runningheads]{llncs}
\usepackage[T1]{fontenc}
\usepackage{graphicx}
\usepackage{hyperref}
\usepackage{color}

\usepackage[misc]{ifsym}

\usepackage{amsmath,amsfonts,bm}

\def\eqref#1{equation~\ref{#1}}

\def\1{\bm{1}}

\def\vb{{\bm{b}}}
\def\vc{{\bm{c}}}

\def\vr{{\bm{r}}}

\def\vv{{\bm{v}}}

\def\mC{{\bm{C}}}

\def\mW{{\bm{W}}}

\DeclareMathAlphabet{\mathsfit}{\encodingdefault}{\sfdefault}{m}{sl}
\SetMathAlphabet{\mathsfit}{bold}{\encodingdefault}{\sfdefault}{bx}{n}

\begin{document}
\title{Amortizing Personalization in Virtual Brain Twins}
%
%
\author{Nina Baldy\inst{1}\orcidID{0009-0000-2444-1837}\Letter \and
Marmaduke M. Woodman\inst{1}\orcidID{0000-0002-8410-4581}\and
Viktor K. Jirsa\inst{1}\orcidID{0000-0002-8251-8860}}
\authorrunning{N. Baldy et al.}
%
\institute{Aix Marseille Univ, INSERM, \\ INS, Inst Neurosci Syst, Marseille, France. \\
\email{marmaduke.woodman@univ-amu.fr}}
\maketitle              

\begin{abstract}
Virtual brain twins are personalized digital models of individual human subject or 
patient's brains, allowing for mechanistic interpretation of neuroimaging data features.
Training \& inference with these models however presents a pair of challenges: (a) large shared
infrastructure do not allow for use of personal data and (b) inference in clinical applications
should not require significant resources.  We introduce "anonymized personalization" to address both by expanding model priors to include
personalization which under amortized inference allows training to be performed anonymously, while
inference is both personalized and lightweight.  
We illustrate the basic approach, demonstrate
reliability in an example, and discuss the impact on both experimental and computational neuroscience. Code is available at \url{https://github.com/ins-amu/apvbt}.
\end{abstract}
\section{Introduction}
\label{intro}

Virtual brain twins (VBTs) represent a synthetic approach in computational neuroscience,
offering personalized digital models of individual brains that enable mechanistic interpretations
of neuroimaging data features ~\cite{hashemi_vbt} \cite{wang2023delineating} \cite{lavanga2023virtual} \cite{hashemi2023amortized} and the underlying causal organization \cite{jirsa2022entropy} \cite{fousek2024symmetry}.
These models hold significant potential for advancing our understanding
of brain function, neurological disorders, and personalized medicine. However, their broad
application present significant challenges that must be addressed to fully realize their clinical
and research utility.

One major challenge lies in the training and inference processes required for these models.
Traditional approaches often rely on large shared infrastructure  \cite{schirner2022brain},
which poses a critical
barrier due to the sensitivity of brain imaging data. The use of such infrastructure requires
stringent privacy protections, complicating the integration of sensitive patient information
into model development. Additionally, use in the clinic requires efficient and resource-light
inference methods to ensure practicality in real-world settings.

To overcome these challenges, we propose expanding model priors to incorporate personalization
within an amortized inference framework. This approach allows for anonymous training while enabling
personalized and lightweight inference, addressing both the privacy concerns associated with
sensitive data and the computational demands of clinical applications. By integrating prior
knowledge about brain structure and function into the model architecture, we can train models
without direct access to individual patient data, thereby preserving privacy while maintaining
the ability to generate accurate and personalized VBTs.

This paper is organized as follows: Section 2 outlines our methodology, detailing how we expand
model priors for personalization and implement amortized inference. Section 3 presents our results,
including an assessment on a test case. Finally, Section 4
discusses the broader implications of this work for
computational neuroscience and clinical applications.

\section{Methods}
\label{methods}

\subsection{Virtual brain twins}
\label{vbtmodels}

\begin{figure}
\begin{center}
   \includegraphics[width=0.8\linewidth]{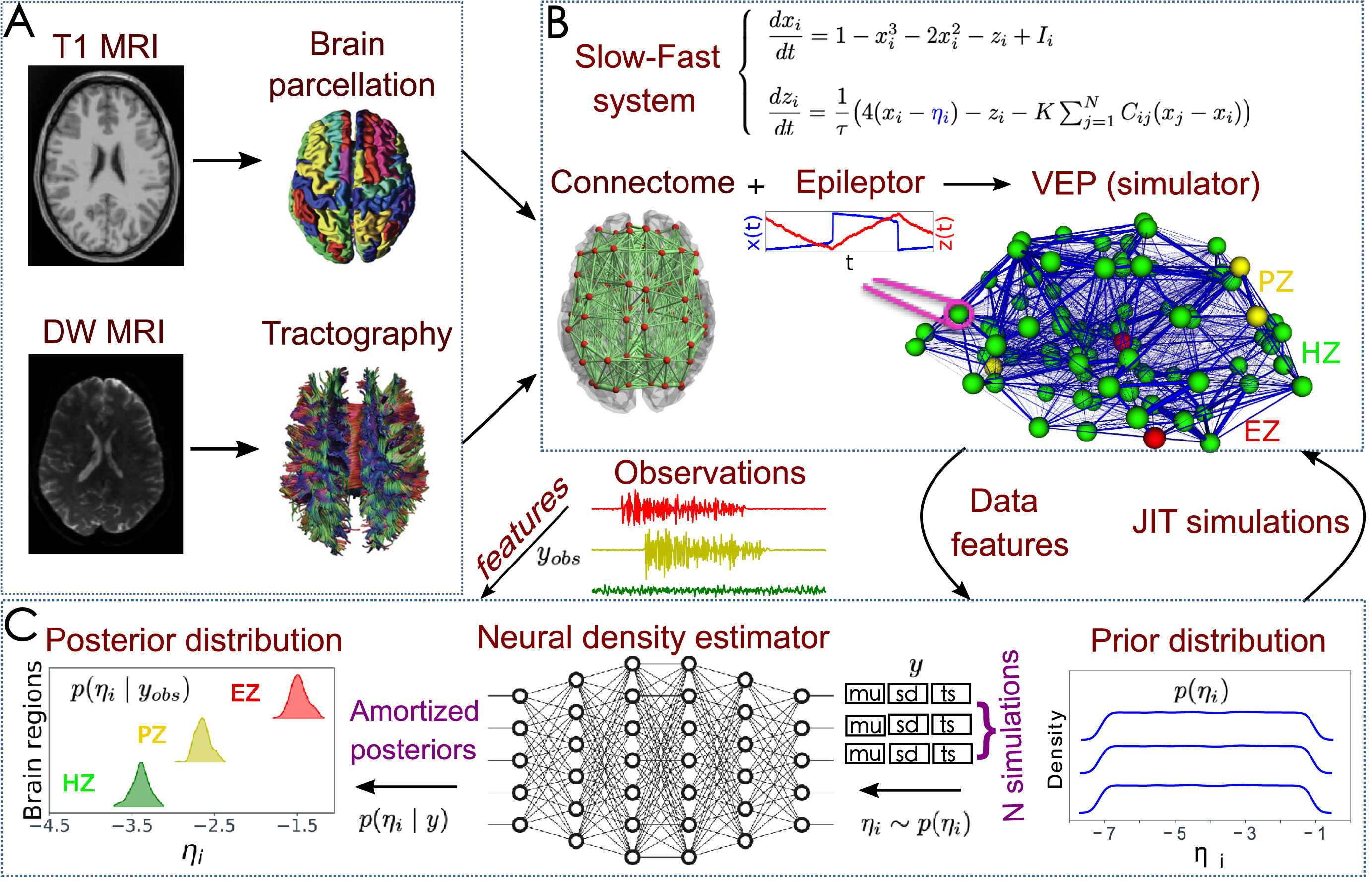}
\end{center}
\caption{
\textbf{Virtual brain twins provide mechanistic interpretation of neuroimaging data.}
\label{sbi-workflow}
An example of amortized inference workflow as developed by Hashemi et al. \cite{hashemi2023amortized} for
clinical epilepsy.  MRI images (T1 \& diffusion-weighted) are used to reconstruct
brain geometry, parcellation \& connectome.  Nonlinear dynamics are assigned to each
node, with connectome providing the interactions between nodes.  The model is simulated
for many samples in parameter space, training a deep Bayesian estimator to map data features to posterior distributions over model parameters.  Once trained, the estimator can be rapidly applied to unseen data to provide new inferences. This figure from Hashemi et al. \cite{hashemi2023amortized} was reproduced with permission.}
\end{figure}

\noindent VBT models describe whole brain neural activity, and, through forward models,
neuroimaging data such as EEG or fMRI.  The models are composed of a set of nodes representing
regions of the brain, each with dynamics described by differential equations, and a connectome,
which describes the weighted connections between the regions,
usually derived from diffusion weighted imaging, as described below in Section \ref{connectomes}.
For the models used here, we
employ a low dimensional description of neural dynamics \cite{montbrio2015macroscopic},
in terms of $N$ nodes' mean firing rates $\vr(t)$
and mean membrane potentials $\vv(t)$,

\begin{align}
    \dot{\vr} &= \Delta/\pi + 2 \vr \vv + k \mC \vr \\
    \dot{\vv} &= \vv^2 + \eta + J \vr + I - \pi^2 \vr^2
\end{align}

where $\mC$ is the matrix of connectome weights, and default parameter values
$\tau=1.0$, $I=0.0$, $\Delta=1.0$, $J=15.0$, $\eta=-5.0$ generate bistable dynamics. Simulations of the VBT model involves sampling parameters of interest,
integrating the differential equations (potentially stochastic and/or delayed), computing relevant data features.
For our tests, we consider as parameters only coupling scaling $k$ and noise scaling $D$  as parameters of interest, with uninformative uniform priors,
i.e. $k\sim U(0.1,0.3)$ and $D~\sim U(0.2,0.4)$.  The data feature used for inference is the mean value of $\vr_i(t)$ at steady state.

A typical workflow for inference with a VBT is illustrated in Figure \ref{sbi-workflow}.
The canonical computational implementation for simulating such models is provided by
The Virtual Brain software package \cite{sanz2013virtual}, freely available from
\url{https://thevirtualbrain.org}.

\subsection{Personal data: connectomes}
\label{connectomes}

In typical use of VBTs, the personal data of interest is the set of 
connection weights describing how regions in the brain communicate,
enabling the model to provide personalized predictions of brain dynamics.
In this report, we make use of MRI scans 461 subjects across a German cohort
in the 1000BRAINS project (1KB) over a range of ages
collected by \cite{caspers2014studying} and an American
cohort of subjects part of the Human Connectome Project (HCP)
described by \cite{van2013wu},
with the connectomes reconstructed under 20
parcellations by \cite{jung2022parc1kb}.
Because the connectomes represent raw counts of tracks between regions with large variance, they are preprocessed by 
(a) clipping values to 1 to 99th percentiles, (a) normalizing between 0 and 1, and (c) 
taking the square root per connectome, then removing the mean over all connectomes, such that
subsequent steps work with residual deviations around preprocessed mean connectomes.

\subsection{Priors over personal data}
\label{priorspersonal}

While inference with VBTs has generally taken the connectome as a fixed piece
of data on which the model and posterior are conditioned, we will instead
 consider the connectome as a parameter to be inferred, alongside the other
parameters of interest. Human connectomes derived from diffusion-weighted imaging
are symmetric, leading to additional $(N^2-N)/2$ parameters (ignoring
self connections).  We regularize this high-dimensional set of parameters by introducing a low-rank
representation, in an approach similar to an autoencoder \cite{hinton1989auto} that we term
\textit{cross-coder} since it collects information across parcellations:
each subject in the
dataset has their connectome represented in 20 parcellations.
We train a two layer architecture to encode the connectome of a subject shown
in Figure \ref{xcode-arch},
from all parcellations into the same latent space, and decode from that latent
space to all parcellations, with a $l_2$ loss:

\begin{figure}
\begin{center}
   \includegraphics[width=0.8\linewidth]{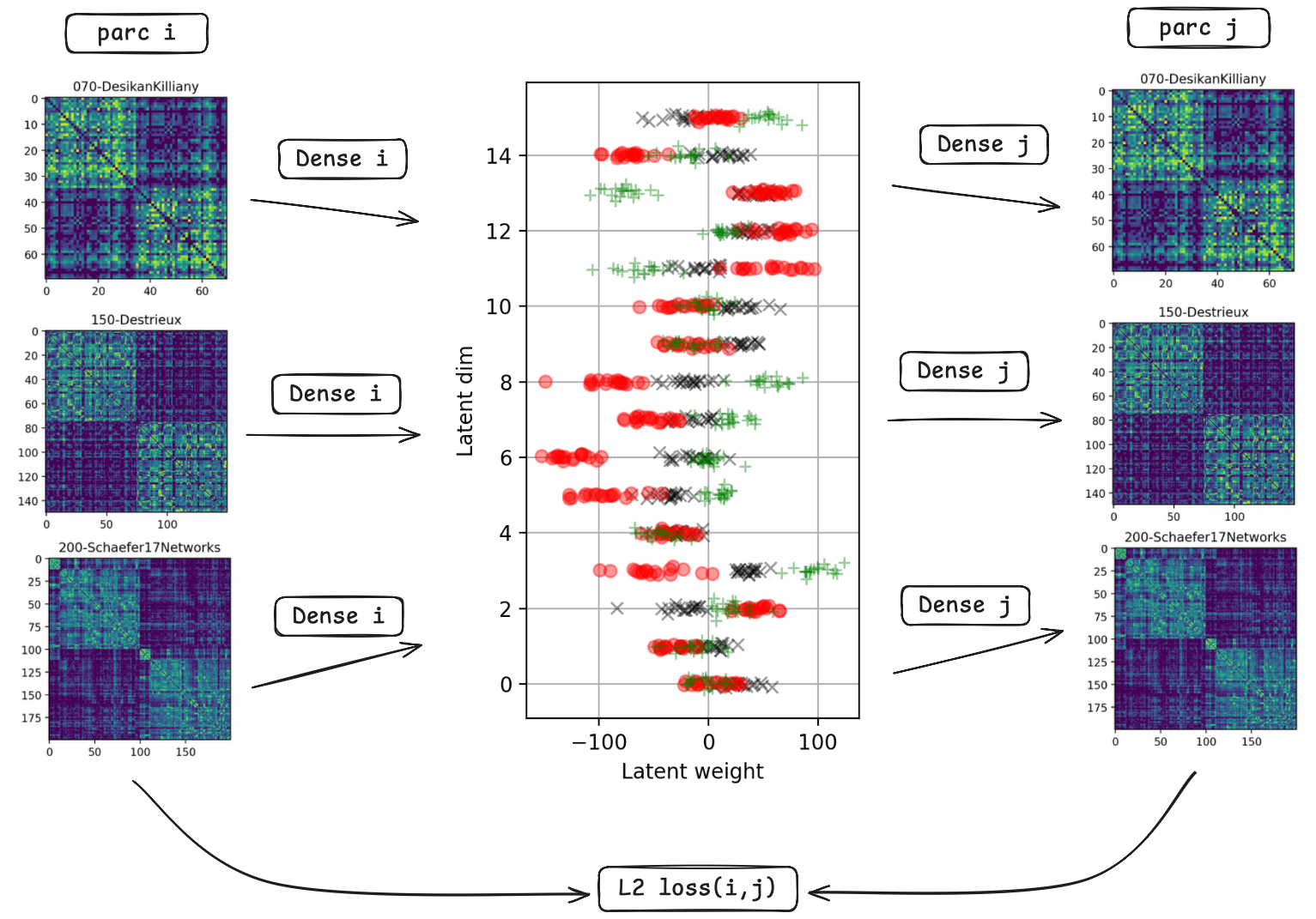}
\end{center}
\caption{
\textbf{Cross-coder architecture encodes connectomes.}
The cross-coder architecture, inspired by autoencoder, takes the subject
connectome data in the $i$th parcellation, encodes with a single
dense layer to a latent space and decodes to the $j$th parcellation,
with a $l_2$ loss on the reconstruction error.  Each encoder is informed
not only by its own parcellation but also errors due to decoding to all
other parcellations, and similarly for each decoder.
} 
\label{xcode-arch}
\end{figure}

\begin{equation}
L_2 = \sum_{i,j,k} \| \vc_{j,k} - \left((\vc_{i,k} \mW_i^e + \vb_i^e)^T \mW_j^d + \vb_j^d\right)^T\|
\end{equation}
where $\vc_{i,k}$ is a vector containing
the upper triangle of the connectome for the $k$th subject in the
$i$th parcellation, $\mW_i^e$ and $\vb_i^e$ are the weights matrices
and bias vectors for the encoder
from the $i$th parcellation, $\mW_i^d$ and $\vb_i^d$ similarly for the decoder. Stochastic gradient descent is performed with the Adam optimizer \cite{kingma2014adam} in JAX \cite{jax},
with a learning rate of 3e-4, until convergence, typically 1000 iterations, with batch size 64, requiring
2 minutes on an RTX 4090 or 20 minutes on Apple M3.
The choice of a simple linear representation ensures (a) fast convergence despite small sample size
(b) robust generalization and (c) inability to learn training data.
A set of encoders \& decoders were trained per cohort, and for each cohort one
half of the connectomes were used for training (train-test split of 0.5).

While the loss function drives the optimization, our application is to personalization,
and we therefore assess the final result in terms of confusion rate, i.e. the probability
of the cross-coder producing a connectome from subject $k$ which is more similar to another
subject $l$ than $k$, with similarity assessed via Euclidean distance.

Finally, to obtain a Bayesian prior, we compute the mean and covariance over the latent
project of all (training) subjects' connectomes from all parcellations.  The resulting mean
and covariance parameterize a multivariate normal distribution usable for sampling.  We remark
that while this distribution is derived from personal data, because it averages over many
subjects, it is no longer personal data subjected to privacy concerns. Such prior distribution of connectivity could be referred to as a cohort-based prior.

\subsection{Simulation based inference}
\label{sbi}

Simulation-based inference (SBI) is a family of techniques that formulates Bayesian
inference as a Bayesian regression problem, in which the data features parameterize
(via a neural network) a distribution over the parameters \cite{cranmer2020frontier}.
The process starts with a training step in which the prior is sampled, and a simulator computes 
corresponding data features for samples from the prior, which then allows training the neural
network. The posterior distribution is thus shaped through a sequence of invertible transformations, and once trained, can be quickly sampled from by providing data features. Here we use amortized neural posterior estimation \cite{papamakarios2016fast} as implemented
by \cite{tejero-cantero2020sbi}
in the PyTorch-based Python package \verb+sbi+, \url{https://sbi-dev.github.io/sbi/latest/}.
We evaluate the results with model-based calibration metrics \cite{betancourt2018calibrating},
specifically posterior shrinkage and posterior z-score

\begin{align}
    s_n &= 1 - \frac{\sigma^2_n(y)}{\tau_n^2(y)} \\
    z_n &= \left| \frac{\mu_n(y) - \theta_n}{\sigma_n(y)} \right|
\end{align}
where $\tau_n$ and $\sigma_n$ are standard deviations of the prior and posterior, $\mu_n$ is the posterior mean and $\theta_n$ is the true parameter value; 
and posterior predictive 90\% confidence intervals. An acceptable inference
process should show high shrinkage, low z-score,
and the true value should lie in the 90\% confidence region.
As the main question studied here is replacing per-subject
SBI (where the connectome is taken as fixed data and not estimated)
by cohort-level SBI (i.e. absorbing the connectome as a parameter with the prior), we
focus on evaluating the degree to which cohort-level SBI matches subject-level SBI
performance.

\section{Results}
\label{results}

\subsection{Cross-coding sufficient for personalized connectomes}
\label{xcodesuff}
Training the cross-coder architecture yields a low rank
representation of connectomes, via gradient descent on a training set of 230 connectomes and 230 connectomes across the 1KB and HCP datasets, for latent dimension of 16. Once validation loss converges,
the confusion rate is 2.3\% on connectomes in the validation set,
confirming that low rank representation successfully preserves the personalization of the connectomes.

Because the cross-coder uses the same latent space for all parcellations, we expected
the different parcellations to project the same subject to similar latent coordinates.
The middle panel in Figure \ref{xcode-arch} shows three example subjects projected to
the latent space of a trained cross-coder, showing much lower intra-subject
variability compared to inter-subject variability.  This verifies a crucial aspect of maintaining
personalization of the connectomes in the low rank latent space.

\subsection{Cohort SBI provides reliable subject-level inference}
\label{subjectlevel}

\begin{figure}
\begin{center}
   \includegraphics[width=0.8\linewidth]{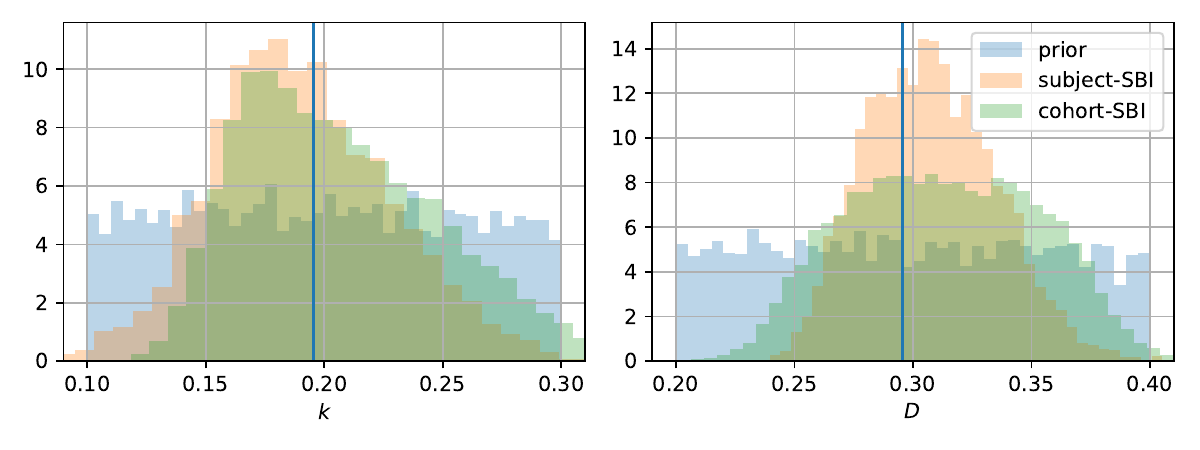}
\end{center}
\caption{
\textbf{Example of subject vs cohort-level inference.}
The prior (blue, uniform) contracts to the posterior
in both subject-level (orange) and cohort-level SBI (green).
The ground truth value is indicated by the blue vertical line.
Both inferences provide similar posteriors.}
\label{kD-subj-cd-dist}
\end{figure}

\noindent We evaluated cohort-level SBI, i.e. SBI over the parameters including the latent
connectome parameters, vs subject-level SBI, where the connectome is fixed. The
results shown are for the 79-node Shen 2013 parcellation only. Figure
\ref{kD-subj-cd-dist} shows the prior distribution and posteriors for the two parameters
of interest: coupling scaling $k$ and noise scaling $D$.  While the posterior obtained with
cohort-level SBI does not contract as significantly as for the subject-level SBI, there
is a clear contraction towards the true value. 

\begin{figure}
\begin{center}
   \includegraphics[width=0.8\linewidth]{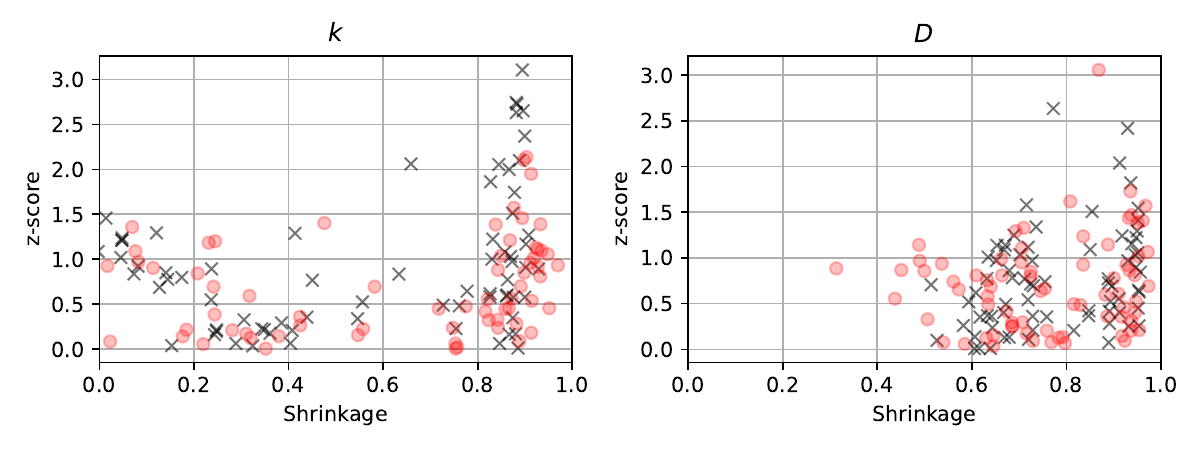}
\end{center}
\caption{
\textbf{Cohort-level SBI approximates subject-level SBI.}
Distributions of model-based calibration scores for all subjects in the
test set (N=230). Black cross represent scores of the cohort-level SBI (applied to
each subject), while red dots indicate scores of each subject-level SBI.}
\label{kD-subj-cd}
\end{figure}

We next sought to evaluate the performance across all subjects in the test set (N=230), by
performing subject-level SBI for each, and comparing with posterior obtained with the
cohort-level SBI.  In Figure \ref{kD-subj-cd}, the metrics reveal a very similar (correlation 0.8) distribution, and the rates at which the true value
lies in the 90\% CI of the posterior are 0.89 and 0.95 for cohort and subject-level SBI respectively. 
These results suggests that the cohort-level SBI provides a useful approximation of subject-level SBI, leveraging an amortized posterior that can be evaluated for any subject - while having been trained only once, with only as many simulations as its subject-level counterpart. The computational complexity of the simulation process, for N subjects and M simulations, is O(NM) for per subject SBI, and O(M) for cohort SBI.

\subsection{Cohort inference identifies the connectome}
\label{idconnectome}

\begin{figure}
\begin{center}
   \includegraphics[width=0.8\linewidth]{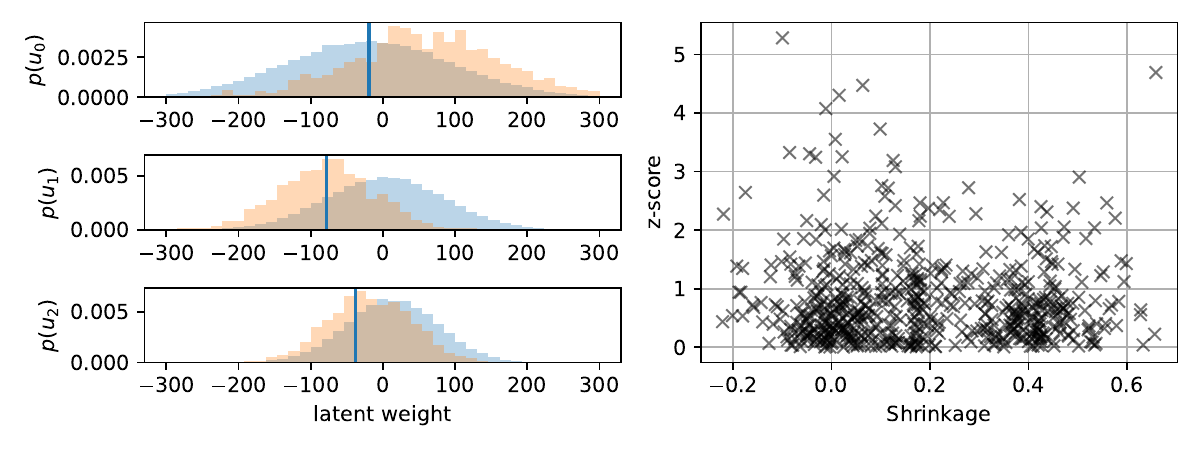}
\end{center}
\caption{
\textbf{Cohort-level SBI partially identifies latent connectome parameters.}
\textit{Left}: The prior (blue) contracts to the posterior
 cohort-level SBI (orange) in most cases, here shown three latent connectome
 parameters. 
 \textit{Right}: Posterior shrinkage \& z-scores for all connectomes in the
 test set, as identified via cohort-level SBI, showing a tendency toward identification
 if not as a strong as for parameters of interest.}
 \label{sbi-conn-id}

\end{figure}

\noindent Lastly, for cohort-level SBI, we sought to evaluate how well the connectome is identified as a parameter.
As shown in Figure \ref{sbi-conn-id}, there is a tendency towards identifying the
latent connectome parameters, but not as strong as for parameters of interest. This may suggest that the model in the regime parameterized for each subject may not always be sensitive to the contribution of the individual connectome. In such case, the average connectome derived from a population could provide a reasonable fit to a given individual. In the more general case, the use of a cohort-based Bayesian prior allows this simplification to take place (through the averaging over the latent project) without excluding outliers that do not match this average.  However, given the metabolic cost of maintaining the connectome \cite{bullmore_sporns}, shrinkage of the posterior
over connectomes may signal a physiologically-relevant regime of the parameters of interest, helping to guide model
construction and evaluation.

\section{Discussion}
\label{discussion}

The development of VBTs with expanded model priors represents a significant
advancement in computational neuroscience. We have demonstrated that incorporating cohort-based
priors over personal data enables amortized inferences on par with per-subject personal inference,
allowing new applications where personal data is unavailable (because it was not recorded or
because of privacy protections). 
While this technique does not replace state of the art, computationally
intensive workflows involving e.g.  MCMC, but rather to a strong compromise between performance,
flexibility and inference efficiency. 

\subsection{Limitations}

The proposed cohort-level inference serves as a useful approximation of subject-specific models but does not fully capture all the individual details of a person's brain connectivity. This is evidenced by the finding that the posterior distributions from cohort-level SBI do not contract as significantly as those from subject-level SBI. Furthermore, while the model parameters of interest like coupling and noise are well-identified, the framework is less successful at strongly identifying the individual latent connectome parameters themselves. Rather than a methodological deficit, we suggest this as degree of model degeneracy where, in certain parameter regimes, the model's dynamics may not be highly sensitive to the fine-grained details of an individual's connectome.

\subsection{Implications for translational research}

This framework significantly lowers the barrier for translating VBTs into clinical and large-scale research settings. The "anonymized personalization" approach allows the computationally expensive training of a model to be performed on large, shared high-performance computing infrastructure without requiring access to sensitive personal data. The resulting trained model can then be used by clinicians for lightweight and personalized inference on new patients without needing complex retraining or patient-specific tuning, enabling practical real-time analysis and decision support. This approach also makes it feasible to scale personalized inferences across thousands of subjects, overcoming the prohibitive computational burden typically associated with large cohorts (computational complexity for N subjects and M simulations is reduced from O(NM) for per-subject SBI to O(M) for cohort SBI).

Our approach also addresses the common research problem of incomplete datasets, enabling new analyses where they were previously impossible: per-subject diffusion imaging needed to create a structural connectome is not always available. This method makes it possible to build and infer parameters for a VBT model in such cases by substituting the missing data with a cohort-based prior derived from a similar population.

\subsection{Future Directions}
\raggedbottom 
A crucial next step is to validate the cohort-based inference framework in specific, well-established domains of whole-brain modeling, namely resting-state dynamics and clinical epilepsy. While this work demonstrates the method's general reliability, applying it to resting-state models would allow for rigorous validation against empirical functional connectivity, a standard benchmark in the field. In epilepsy, where personalized models are already used to delineate epileptogenic networks, this approach could significantly accelerate clinical translation by enabling models to be trained on large, anonymized cohorts and then rapidly applied to new patients, testing the framework's ability to capture both normative and pathological brain states.

A second extension exploiting amortization lies in high-resolution brain models.  The brain models employed in this study are based on "low resolution" connectomes with tens to hundreds of regions. Recent work \cite{wang2023delineating} clearly motivated the need for spatially detailed models for resolving causal effects, however these models require even more resources to run.  To ensure inference
is tractable for a wide range of users, our amortized approach should be evaluated.

Finally, the computational efficiency of the amortized inference process makes this approach a prime candidate for novel real-time clinical applications, such as model-based neurofeedback. Standard inference methods are too slow for real-world clinical use , but once a cohort model is trained, the amortized posterior can be evaluated almost instantly for a new subject. This speed is essential for neurofeedback, where a subject's brain state must be inferred and fed back to them with minimal delay. A VBT operating in real-time could offer mechanistic insights to help guide a patient in modulating their own brain activity, representing a powerful therapeutic avenue that is infeasible with current computationally intensive techniques.

\subsection{Conclusions}

Amortizing personalization in VBTs represents a significant step toward making sophisticated, mechanistic brain models practical tools for both science and medicine. By expanding model priors to include cohort-level information, our method successfully enables anonymous training and lightweight, efficient inference while still retaining individual personalization. This approach provides a strong compromise between performance and efficiency, overcoming critical barriers related to data privacy and computational cost that have hindered the widespread adoption of VBTs. Ultimately, this work helps bridge the gap between complex research models and actionable insights in real-world clinical environments.

\begin{credits}
\subsubsection{\ackname} 
This project/research has received funding from the European Union’s Horizon Europe Programme under the Specific Grant Agreements No. 101147319 (EBRAINS 2.0 Project) and No. 101137289 (Virtual Brain Twin Project).

\subsubsection{\discintname}
The authors have no competing interests to declare that are
relevant to the content of this article. 
\end{credits}
%
%
%
\bibliographystyle{splncs04}
\bibliography{ref}

\begin{thebibliography}{10}
\providecommand{\url}[1]{\texttt{#1}}
\providecommand{\urlprefix}{URL }
\providecommand{\doi}[1]{https://doi.org/#1}

\bibitem{betancourt2018calibrating}
Betancourt, M.: Calibrating model-based inferences and decisions. arXiv
  preprint arXiv:1803.08393  (2018)

\bibitem{bullmore_sporns}
Bullmore, E., Sporns, o.: The economy of brain network organization. Nature
  Reviews Neuroscience  \textbf{13},  336–349 (2012). \doi{10.1038/nrn3214}

\bibitem{caspers2014studying}
Caspers, S., Moebus, S., Lux, S., Pundt, N., Sch{\"u}tz, H., M{\"u}hleisen,
  T.W., Gras, V., Eickhoff, S.B., Romanzetti, S., St{\"o}cker, T., et~al.:
  Studying variability in human brain aging in a population-based german
  cohort—rationale and design of 1000brains. Frontiers in aging neuroscience
  \textbf{6}, ~149 (2014)

\bibitem{cranmer2020frontier}
Cranmer, K., Brehmer, J., Louppe, G.: The frontier of simulation-based
  inference. Proceedings of the National Academy of Sciences  \textbf{117}(48),
   30055--30062 (2020)

\bibitem{fousek2024symmetry}
Fousek, J., Rabuffo, G., Gudibanda, K., Sheheitli, H., Petkoski, S., Jirsa, V.:
  Symmetry breaking organizes the brain’s resting state manifold. Scientific
  reports  \textbf{14}(1),  31970 (2024)

\bibitem{hashemi_vbt}
Hashemi, M., Depannemaecker, D., Saggio, M., Triebkorn, P., Rabuffo, G.,
  Fousek, J., Ziaeemehr, A., Sip, V., Athanasiadis, A., Breyton, M., Woodman,
  M., Wang, H., Petkoski, S., Sorrentino, P., Jirsa, V.: Principles and
  operation of virtual brain twins. IEEE Reviews in Biomedical Engineering pp.
  1--25 (2025). \doi{10.1109/RBME.2025.3562951}

\bibitem{hashemi2023amortized}
Hashemi, M., Vattikonda, A.N., Jha, J., Sip, V., Woodman, M.M., Bartolomei, F.,
  Jirsa, V.K.: Amortized bayesian inference on generative dynamical network
  models of epilepsy using deep neural density estimators. Neural Networks
  \textbf{163},  178--194 (2023)

\bibitem{hinton1989auto}
Hinton, G.E.: Connectionist learning procedures. Artificial Intelligence
  \textbf{40}(1),  185--234 (1989).
  \doi{https://doi.org/10.1016/0004-3702(89)90049-0}

\bibitem{jax}
JAX-authors: Jax: Composable transformations of python+numpy programs (2025),
  \url{https://github.com/google/jax}, accessed: 2025-02-06

\bibitem{jirsa2022entropy}
Jirsa, V., Sheheitli, H.: Entropy, free energy, symmetry and dynamics in the
  brain. Journal of Physics: Complexity  \textbf{3}(1),  015007 (2022)

\bibitem{kingma2014adam}
Kingma, D.P., Ba, J.: Adam: A method for stochastic optimization. arXiv
  preprint arXiv:1412.6980  (2014)

\bibitem{jung2022parc1kb}
Kyesam, J., Eickhoff, S.B., Popovych, O.V.: Parcellation-based structural and
  resting-state functional whole-brain connectomes of 1000brains cohort (v1.1)
  (2022)

\bibitem{lavanga2023virtual}
Lavanga, M., Stumme, J., Yalcinkaya, B.H., Fousek, J., Jockwitz, C., Sheheitli,
  H., Bittner, N., Hashemi, M., Petkoski, S., Caspers, S., et~al.: The virtual
  aging brain: Causal inference supports interhemispheric dedifferentiation in
  healthy aging. NeuroImage  \textbf{283},  120403 (2023)

\bibitem{montbrio2015macroscopic}
Montbri{\'o}, E., Paz{\'o}, D., Roxin, A.: Macroscopic description for networks
  of spiking neurons. Physical Review X  \textbf{5}(2),  021028 (2015)

\bibitem{papamakarios2016fast}
Papamakarios, G., Murray, I.: Fast $\varepsilon$-free inference of simulation
  models with bayesian conditional density estimation. Advances in neural
  information processing systems  \textbf{29} (2016)

\bibitem{sanz2013virtual}
Sanz~Leon, P., Knock, S.A., Woodman, M.M., Domide, L., Mersmann, J., McIntosh,
  A.R., Jirsa, V.: The virtual brain: a simulator of primate brain network
  dynamics. Frontiers in neuroinformatics  \textbf{7}, ~10 (2013)

\bibitem{schirner2022brain}
Schirner, M., Domide, L., Perdikis, D., Triebkorn, P., Stefanovski, L., Pai,
  R., Prodan, P., Valean, B., Palmer, J., Langford, C., et~al.: Brain
  simulation as a cloud service: The virtual brain on ebrains. NeuroImage
  \textbf{251},  118973 (2022)

\bibitem{tejero-cantero2020sbi}
Tejero-Cantero, A., Boelts, J., Deistler, M., Lueckmann, J.M., Durkan, C.,
  Gonçalves, P.J., Greenberg, D.S., Macke, J.H.: sbi: A toolkit for
  simulation-based inference. Journal of Open Source Software  \textbf{5}(52),
  ~2505 (2020). \doi{10.21105/joss.02505},
  \url{https://doi.org/10.21105/joss.02505}

\bibitem{van2013wu}
Van~Essen, D.C., Smith, S.M., Barch, D.M., Behrens, T.E., Yacoub, E., Ugurbil,
  K., Consortium, W.M.H., et~al.: The wu-minn human connectome project: an
  overview. Neuroimage  \textbf{80},  62--79 (2013)

\bibitem{wang2023delineating}
Wang, H.E., Woodman, M., Triebkorn, P., Lemarechal, J.D., Jha, J., Dollomaja,
  B., Vattikonda, A.N., Sip, V., Medina~Villalon, S., Hashemi, M., et~al.:
  Delineating epileptogenic networks using brain imaging data and personalized
  modeling in drug-resistant epilepsy. Science Translational Medicine
  \textbf{15}(680),  eabp8982 (2023)

\end{thebibliography}

\end{document}